\documentclass[pra,aps,10pt,twocolumn]{revtex4-1}
\usepackage{bm,amssymb,graphicx}

\newcommand{\om}{\omega}

\newcommand{\bk}{{\bf k}}
\newcommand{\ep}{\varepsilon}

\begin{document}
\title{Engineering arbitrary pure and mixed quantum states}
\author{Alexander Pechen}
\email{apechen@gmail.com}
\homepage{http://www.mathnet.ru/eng/person17991}
\affiliation{Steklov Mathematical Institute of Russian Academy of Sciences,
Gubkina str. 8, Moscow 119991, Russia\\ Department of Chemical Physics, Weizmann
Institute of Science, Rehovot 76100, Israel}

\begin{abstract}This work addresses a
fundamental problem of controllability of open quantum systems, meaning the
ability to steer arbitrary initial system density matrix into any final density
matrix. We show that under certain general
conditions open quantum systems are completely controllable and propose the
first, to the best of our knowledge, deterministic method for a
laboratory realization of such controllability which allows for a practical
engineering of arbitrary pure and mixed quantum states. The method exploits
manipulation by the system with a laser field and a
tailored nonequilibrium and time-dependent state of the surrounding
environment. As a physical example of the environment we consider incoherent
light, where control is its spectral density.
The method has two specifically important properties: it realizes the strongest
possible degree of quantum state
control --- complete density matrix controllability which is the ability
to steer
arbitrary pure and mixed initial states into any desired pure or mixed final
state, and is ``all-to-one'', i.e. such that each particular control can
transfer simultaneously all initial
system states into one target state.
\end{abstract}
\maketitle

Controlled manipulation by atoms and molecules using external controls is an
active field of modern research with applications ranging from selective
creation of atomic or molecular excitations out to control of chemical reactions
or design of nanoscale systems with desired properties. The external control may
be either coherent (e.g., a tailored laser pulse~\cite{Ra88-90}) or incoherent
(e.g., a specially adjusted or engineered environment~\cite{PeRa06,ICE} or
quantum measurements commonly used with and sometimes without
feedback~\cite{QMC}).

A challenging topic in quantum control is to provide practical methods for
engineering arbitrary quantum states~\cite{Eberly}. The interest to this topic
is driven by fundamental connections to quantum physics as well as by potential
applications to quantum state measurement~\cite{Eberly} and quantum computing
with mixed states and non-unitary quantum gates~\cite{Tarasov2002}. Various
recipes for engineering arbitrary quantum states of light were
proposed~\cite{Vogel1993,Bimbard2010}. For matter, engineering arbitrary open
system's quantum dynamics with coherent control, quantum measurements and
feedback was shown to be achievable~\cite{Lloyd2001}. However, the problem of
deterministic engineering of arbitrary quantum states of matter has generally
been remained unsolved.

This works proposes the first, to the best of the author's knowledge,
deterministic method for engineering arbitrary pure and mixed density matrices
for a wide class of quantum systems. The method uses a combination of incoherent
control by engineering the state of the environment (for which we consider
appropriately filtered incoherent radiation) on the time scale of several orders
of magnitude of the characteristic system relaxation time $\tau_{\rm rel}$
followed by fast (e.g., femto-second) coherent laser control to produce
arbitrary pure and mixed states for a wide class of quantum systems. The method
is deterministic in the sense that it does not use real-time feedback and can be
applied to an ensemble of systems without the need for an individual addressing
of each system. Important is that the suggested scheme requires the
ability to manipulate by both Hamiltonian and non-Hamiltonian aspects of the
dynamics; a fundamental result of Altafini shows that varying only the
Hamiltonian is not sufficient to produce arbitrary states of a quantum
system~\cite{Altafini}.

Engineered environments were suggested for improving quantum computation and
quantum state engineering~\cite{Cirac2009}, making robust quantum
memories~\cite{Cirac2011}, preparing many-body states and non-equilibrium
quantum phases~\cite{Zoller2008}, inducing multiparticle entanglement
dynamics~\cite{Blatt2010}. This work exploits incoherent control by engineered
environment~\cite{PeRa06} in a combination with coherent control for producing
arbitrary pure and mixed density matrices. While incoherent
processes were used in various circumstances, as for example in cold molecule
research, to prepare a pure state needed for a full control over the system's
pure states~\cite{Lasercooling}, their use in the proposed method serves for a
more general goal of engineering arbitrary pure and mixed quantum states.

The method has two special properties. First, it implements complete density
matrix controllability --- the strongest possible degree of state control for
quantum systems meaning the ability to prepare in a controllable way any density
matrix starting from any initial state. Second, the produced controls are
all-to-one --- any such control $c_*$ transfers all pure and mixed initial
states into the same final state and thus can be optimal simultaneously for all
initial states~\cite{Wu07}. This property has no analog for purely coherent
control of closed quantum systems with unitary evolution, where different
initial states in general require different optimal controls. While an abstract
theoretical construction of all-to-one controls was provided~\cite{Wu07}, the
problem of their physical realization has remained open. The suggested method
provides a solution for such a physical realization.

\section{Coherent and incoherent controls} The dynamics of a controlled
$n$-level quantum system isolated from the environment is described by density
matrix $\rho_t$ satisfying the equation
\begin{equation}\label{Sec2:eq2}
\dot\rho_t=-i\left[H_0+u(t)V,\rho_t\right],\qquad \rho_{t=0}=\rho_{\rm i}\, .
\end{equation}
Here $H_0=\sum\ep_i |i\rangle\langle i|$ is the free system Hamiltonian (with
eigenvalues $\ep_1<\ep_2<\dots<\ep_n$ and eigenvectors $|i\rangle$) and $V$ is
the interaction Hamiltonian describing coupling of the system to the control
field $u(t)$ (e.g., a shaped laser pulse)l; commonly $V=-\mu$, where $\mu$ is
the dipole moment of the system. The evolution is unitary,
$\rho_t=U_t\rho_0 U^\dagger_t$, where the unitary evolution operator $U_t$
satisfies the Schr\"odinger equation $\dot U_t=-i\left[H_0+u(t)V\right] U_t$.
The unitary nature of the evolution induced by the field $u(t)$ implies
preservation of coherence in the system such that for example pure states will
always remain pure; the corresponding control is called coherent.

If the system interacts with the environment, then its evolution in the absence
of coherent control ($u(t)=0$) is described by the master equation for the
reduced density matrix
$\rho_t$~\cite{BreuerBook,TarasovBook}
\begin{equation}\label{eq1} \dot\rho_t=-i\left[H_0+H_{\rm
eff},\rho_t\right]+{\cal L}(\rho_t), \qquad \rho_{t=0}=\rho_{\rm i}
\end{equation}
where the superoperator $\cal L$ and the effective Hamiltonian $H_{\rm eff}$
describe the influence of the environment. The effective Hamiltonian $H_{\rm
eff}$ represents spectral broadening of the system energy levels and typically
commutes with $H_0$. For a Markovian environment the superoperator $\cal L$ has
the form ${\cal L}(\rho)=\sum\limits_i(2 L_i\rho L^\dagger_i-L^\dagger_i
L_i\rho-\rho L^\dagger_i L_i)$ with some matrices
$L_i$~\cite{BreuerBook,Lindblad}. The explicit form of these matrices is
determined by the state of the environment and by the details of the microscopic
interaction between the system and the environment.

The matrices $L_i$ are usually considered as fixed and having deleterious effect
on the ability to control the system --- open quantum systems subject to the
Markovian evolution~(\ref{eq1}) with constant $\cal L$ are
uncontrollable~\cite{Altafini}. However, the assumption of constant $\cal L$ is
too restrictive, since $\cal L$ can be manipulated by adjusting the state of the
environment through its temperature, pressure, or more generally through its
distribution function (spectral density). Control through adjusting the state of
the environment in
general does not preserve quantum coherence in the controlled system and for
this reason is called incoherent~\cite{ICE}.

This work considers incoherent radiation as an example of a Markovian control
environment and the scheme is analyzed below for this case. Other environments,
either Markovian or non-Markovian, can also be used as incoherent controls;
however, the ability to use a particular environment for engineering arbitrary
quantum states using the proposed scheme requires a separate
analysis in each case. The state of the environment formed by incoherent photons
is characterized at a time moment $t$ by spectral density $n_\bk(t)$ of
photons with momentum $\bk$; in
general, spectral density of photons with polarization $\alpha$ can also be
exploited.
For the purpose of this work the directional dependence of spectral density is
not necessary and its dependence only in the photon energy
$\om=|\bk|$ is used; $n_\om$ is assumed to be constant over the frequency range
of significant absorption and emission for each system transition frequency
$\om_{ij}=\ep_j-\ep_i$. The spectral density $n_\om(t)$ can be experimentally
manipulated for example by filtering.

The evolution of the system in the environment formed by incoherent radiation
with spectral density $n_\om$ is described by the master
equation~(\ref{eq1}) with superoperator ${\cal L}={\cal L}_{n_\om}$ of the
form~\cite{DaviesBook,SpohnReview}
\begin{equation}
{\cal L}_{n_\omega}(\rho)=\sum\limits_{i<j} A_{ij}
\left[(n_{\omega_{ij}}+1)L_{Q_{ij}}(\rho)+n_{\omega_{ij}}L_{Q_{ji}}
(\rho)\right ]\label{MEWCL}
\end{equation}
where $A_{ij}\ge 0$ are the Einstein coefficients for spontaneous emission,
$\omega_{ij}=\varepsilon_i-\varepsilon_j$ are the system transition frequencies,
$Q_{ij}=|i\rangle\langle j|$ is the transition operator for the
$|j\rangle\to|i\rangle$ transition, and $L_Q(\rho)=2Q\rho Q^\dagger-Q^\dagger
Q\rho-\rho Q^\dagger Q$. The intensities $u(t)$ and $n_\omega$ of the
coherent and incoherent light are the coherent and incoherent controls,
respectively. Radiative energy density per unit angular frequency interval is
$\rho_\om=\hbar\om n_\om
(\om^2/\pi^2c^3)$.

\section{Engineering arbitrary quantum states} Let $\rho_{\rm i}$ be any (mixed
or pure) initial state of the system and $\rho_{\rm f}=\sum p_i
|\phi_i\rangle\langle\phi_i|$ be an arbitrary (mixed or pure) target (final)
state. Without loss of generality we assume
$p_1\ge p_2\ge\dots\ge p_n$. The
system is assumed to be generic in the sense that all its transition frequencies
are different (in particular, its spectrum is non-degenerate) and all
Einstein coefficients $A_{ij}$ are non-zero. The
goal is to find a combination of coherent and incoherent fields transforming all
$\rho_{\rm i}$ into $\rho_{\rm f}$.

The scheme to steer $\rho_{\rm i}$ into $\rho_{\rm f}$ consists of two stages.
In the first stage, the system evolves on the time scale of several orders of
magnitude of the relaxation time, $t\approx a \tau_{\rm rel}$ (where $a$ can be
chosen in the range $2\div 10$ depending on the required degree of accuracy)
under the action of a suitable optimal incoherent control $n^*_\om$ into the
state $\tilde\rho_{\rm f}=\sum p_i|i\rangle\langle i|$ diagonal in the basis of
$H_0$ and having the same spectrum as the final state $\rho_{\rm f}$. The state
$\tilde\rho_{\rm f}$ has the same purity as $\rho_{\rm f}$; $\tilde\rho_{\rm f}$
is mixed if $\rho_{\rm f}$ is mixed and $\tilde\rho_{\rm f}$ is pure if
$\rho_{\rm f}$ is pure. In the second stage, the system evolves on the fast
(e.g. femto-second) time scale under the action of a suitable coherent laser
control which rotates the basis of $H_0$ to match the basis of $\rho_{\rm f}$.

The first stage exploits incoherent control with any $n^*_\om$ such that
$n^*_{\om_{ij}}=p_j/(p_i-p_j)$ to prepare the system in the state
$\tilde\rho_{\rm f}$. Coherent control is switched off ($u(t)=0$) during this
stage and the system dynamics is described by the master equation~(\ref{eq1})
which for off-diagonal matrix elements $\rho_{ln}=\langle l|\rho|n\rangle$ of
the density matrix takes the form
\[
 \dot\rho_{ln}=-(i\alpha_{ln}+\Gamma_{ln})\rho_{ln},\qquad
\Gamma_{ln}=\sum_j(W_{jl}+W_{jn})\ge 0
\]
where $\alpha_{ln}=\langle l|H_0+H_{\rm eff}|l\rangle-\langle n|H_0+H_{\rm
eff}|n\rangle$ and
$W_{ij}=A_{ij}(2n_{\omega_{ij}}+1)$.
Off-diagonal
elements decay exponentially for $\Gamma_{ln}>0$, $\rho_{ln}\sim
\exp(-\Gamma_{ln}t)$. Diagonal elements satisfy the Pauli master equation
\[
 \dot\rho_{nn}=\sum\limits_j (W_{nj}\rho_{jj}-W_{jn}\rho_{nn})
\]
For generic systems, all $n_{\om_{ij}}$ can be independently adjusted, and the
detailed balance condition $W_{nj}\rho_{jj}=W_{jn}\rho_{nn}$ implies
$n^*_{\om_{ij}}=p_j/(p_i-p_j)$. The master equation for spectral density
$n_\om$ such that $n^*_{\om_{ij}}=p_j/(p_i-p_j)$ has
$\tilde\rho_{\rm f}$ as the stationary state and exponentially fast drives the
system to $\tilde\rho_{\rm f}$~\cite{Accardi06}. The case when some $p_i=p_j\ne
0$ formally requires infinite density at the corresponding transition (e.g.
infinite temperature of the environment is required for creating equally
populated state $\tilde\rho_{\rm f}=\frac{1}{n}\mathbb I$), but for practical
applications a reasonable degree of accuracy allowing for a finite
$n^*_{\om_{ij}}$ is always sufficient. The first stage when necessary can be
divided into two parts, diagonalization of the density matrix with any $n_\om$
producing positive and sufficiently large $\Gamma_{ln}$, followed by the
evolution produced by any $n^*_\om$ such that $n^*_{\om_{ij}}=p_j/(p_i-p_j)$ to
produce $\tilde\rho_{\rm f}$.

The second stage implements coherent dynamics with unitary evolution
transforming the basis $\{|i\rangle\}$ into $\{|\phi_i\rangle\}$ and hence
steering $\tilde\rho_{\rm f}$ into $\rho_{\rm f}$. For successful realization
of this stage we assume that any unitary evolution operator of the system can be
produced with available coherent controls when ${\cal L}=0$ on a time scale
sufficiently shorter than $\tau_{\rm rel}$, i.e. that the system is unitary
controllable when decoherence effects are negligible. Incoherent control is
switched off during this stage by setting $n_\om=0$ and the dynamics is
well approximated by the unitary evolution~(\ref{Sec2:eq2}).

Necessary and sufficient conditions for unitary controllability were obtained by
Ramakrishna et al.~\cite{Ramakrishna}. Sufficient conditions for complete
controllability of $n$-level quantum systems subject to a single control pulse
that addresses multiple allowed transitions concurrently were
established~\cite{Schirmer2001}. Ramakrishna et al. showed that a necessary and
sufficient condition for unitary controllability of a quantum system with
Hamiltonian $H=H_0+u(t)V$ is that the Lie algebra generated by the operators
$H_0$ and $V$ has dimension $n^2$. We assume that the system satisfies this
condition when ${\cal L}=0$ and that any $U$ can be produced on a time scale
sufficiently shorter than the decoherence time scale, e.g. by using time-optimal
control~\cite{Khaneja2001}. Under these assumptions a coherent field $u^*(t)$
exists which produces a unitary operator $U$ transforming the basis
$\{|i\rangle\}$ into $\{|\phi_i\rangle\}$ and therefore steering
$\tilde\rho_{\rm f}$ into $\rho_{\rm f}$. This field implements the second stage
of the scheme to finish evolving $\rho_{\rm i}$ into $\rho_{\rm f}$.  The second
stage can be implemented also on the long time scale using dynamical
decoupling~\cite{DD} to effectively decouple the system from the environment.

There exist other methods to prepare arbitrary quantum states. An example is the
method of engineering arbitrary Kraus maps proposed by Lloyd and
Viola~\cite{Lloyd2001} which in particular can be used to produce arbitrary
quantum states. Another option is to cool an ensemble of systems to a pure state
$|\phi\rangle$ and then apply randomly and independently to each system a
control producing a unitary operator $U_i$ transforming $|\phi\rangle$ into
$|\phi_i\rangle$. Implementing each $U_i$ with probability $p_i$ will
effectively produce the systems in the state $\rho_{\rm f}=\sum
p_i|\phi_i\rangle\langle\phi_i|$. Both these methods require independent
addressing of individual systems in the ensemble that in general is hard to
realize. The proposed method is free of this shortcoming and can be applied to
an ensemble without the need to independently address each system. This
advantage should significantly simplify practical realization when individual
addressing of quantum systems is hard to realize. We remark that the first
control stage requires individual addressing of each transition frequency. If
the system is non-generic, in particular, is degenerate then the method may work
if the degenerate states can be selectively addressed by using polarization of
the incoherent radiation. For degenerate systems (harmonic oscillator) another
scheme was proposed~\cite{Eberly}.

Determining the correct form of the master equation for systems with time
dependent Hamiltonians in general is a non-trivial
problem~\cite{arxiv1010.0940,Alickiprivate}. However, the proposed scheme does
not use the master equation with time dependent Hamiltonian $u(t)V$ and
therefore the problem of choosing the correct form of the master equation with
time dependent Hamiltonian is not relevant for this work. The scheme in its
first step exploits incoherent control when the coherent control is switched off
and therefore the system Hamiltonian is time independent. In this case the
system dynamics under Markovian approximation is governed by a master
equation of the form~(\ref{eq1}). The second step exploits fast coherent control
on a short time scale when incoherent control is switched off and the
decoherence effects are negligible. In this case ${\cal L}\approx 0$ and the
dynamics is well approximated by~(\ref{Sec2:eq2}).

\begin{figure}[h]
\includegraphics[scale=0.53]{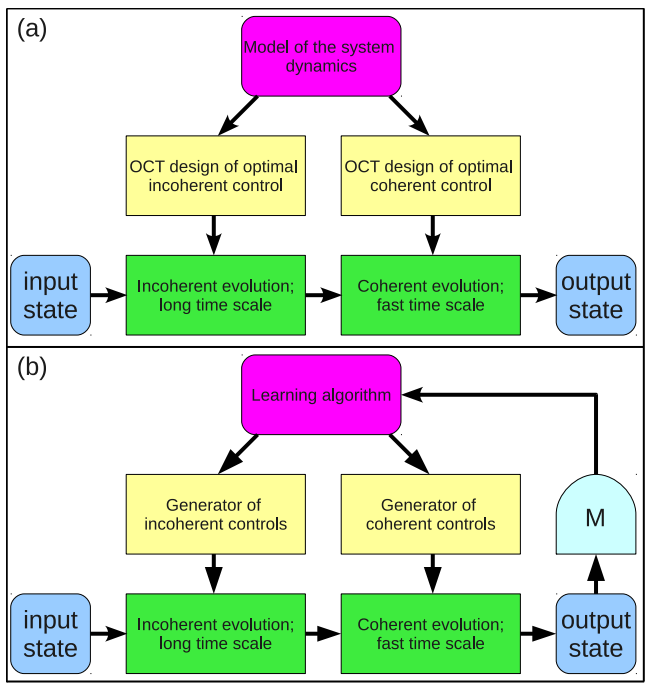}
\caption{(Color online). Implementing complete density matrix controllability
with (a) optimal control theory (OCT) and (b) learning control. In (a), methods
of OCT and a theoretical model of the system dynamics are used to find optimal
incoherent and coherent controls $n^*_k$ and $u^*(t)$. In (b), the objective
$J=\|\rho_T-\rho_{\rm f}\|$, where $\rho_T$ is the output state, is measured
(block with letter ``M``) at the end of each iteration and the result is used to
design a new control, ideally with better performance. The cycle is repeated
until obtaining a satisfactory output.}
\label{scheme}
\end{figure}

If the quantum system is sufficiently simple and all its relevant parameters are
known, then methods of optimal control theory (OCT) can be used to find optimal
fields $n^*_\omega$ and $u^*(t)$ as shown on Fig.~\ref{scheme}, subplot (a). If
the system is complex and (or) some of its relevant parameters are unknown then
various adaptive learning algorithms can be used to implement the proposed
scheme for engineering arbitrary quantum states (Fig.~\ref{scheme}, subplot
(b)). In this case, the creation of a desired diagonal mixed state
$\tilde\rho_{\rm f}$ during the first stage of the control scheme can be
realized with learning algorithms~\cite{PeRa06}. The second stage can be
implemented using learning algorithms for coherent control~\cite{JuRa92}.
Learning algorithms were shown to be efficient for both cases and therefore they
will be efficient when used in a successive combination as required in the
proposed scheme.

\section{Complete density matrix controllability} Depending on a particular
problem, various notions of controllability for quantum systems are
used~\cite{Altafini,Ramakrishna,Schirmer2002,Wu07,Brumer2008,Polack2009},
including unitary controllability, pure state, density matrix and observable
controllability~\cite{Schirmer2002}, and complete density matrix
controllability~\cite{Wu07}. Unitary controllability means the ability to
produce with available controls any unitary evolution operator $U$. Density
matrix controllability means the ability to transfer one into another arbitrary
density matrices with the same spectrum (i.e., kinematically equivalent density
matrices); a particular case is pure state controllability as the ability to
transfer one into another arbitrary pure states. Complete density matrix
controllability means the ability to steer any initial (pure or mixed) density
matrix $\rho_{\rm i}$ into any (pure or mixed) final density matrix $\rho_{\rm
f}$, irrespective of their relative  spectra~\cite{Wu07}. This notion is
different from density matrix controllability~\cite{Schirmer2002}, where only
kinematically equivalent density matrices are required to be accessible one from
another, and is the strongest among all degrees of state controllability for
quantum systems; complete density matrix controllability of a quantum system
implies in particular its pure state and density matrix controllability. The
suggested scheme allows for transferring one into another arbitrary density
matrices thereby approximately realizing complete density matrix controllability
of quantum systems --- the strongest possible degree of their state control.

\section{All-to-one controls} An all-to-one control $c_*$ is a control which
steers all initial states into one final state. Such controls can be optimal
simultaneously for all initial states~\cite{Wu07} and their importance is
motivated by the following. Let $J(c)\equiv J(\rho_T)$ be an arbitrary control
objective determined by the system density matrix $\rho_T$ evolving from the
initial state $\rho_0$ to the final time $T$ under the action of the control $c$
(e.g., $J={\rm Tr}\,[\rho_TO ]$ for some Hermitian observable $O$ or
$J=\|\rho_T-\rho_{\rm f}\|$). In general, optimal controls --- controls
minimizing the objective --- are different for different initial states.
However, if $c_*$ is an all-to-one control steering all states into a state
$\rho_{\rm f}$ minimizing $J$, then $c_*$ will optimize the objective for any
initial system state. While an abstract theoretical construction of all-to-one
controls was provided~\cite{Wu07} in terms of special Kraus maps whose
definition is also provided below, their physical realizations has remained as
an open problem. The proposed control scheme provides a physical realization of
all-to-one controls and the corresponding Kraus maps for any pure or mixed state
$\rho_{\rm f}$. The all-to-one property is achieved during the first stage,
where $n^*_\om$ produces the same density matrix $\tilde\rho_{\rm f}$
independently of the initial state.

The density matrix representing the state of an $n$-level quantum system is a
positive, unit-trace $n \times n$ matrix. We denote by $\mathcal{M}_n
=\mathbb{C}^{n\times n}$ the set of all $n \times n$ complex matrices, and by
$\mathcal{D}_n := \{ \rho\in \mathcal{M}_n \,|\, \rho=\rho^\dagger, \rho \ge 0,
\mathrm{Tr}(\rho) = 1\}$ the set of all density matrices. A map $\Phi :
\mathcal{M}_n \to \mathcal{M}_n$ is positive if $\Phi(\rho)\ge 0$ for any $\rho
\ge 0$ in $\mathcal{M}_n$. A map $\Phi : \mathcal{M}_n
\to \mathcal{M}_n$ is completely positive (CP) if for any $l \in \mathbb{N}$ the
map $\Phi \otimes \mathbb{I}_l : \mathcal{M}_n \otimes\mathcal{M}_l \to
\mathcal{M}_n \otimes \mathcal{M}_l$ is positive ($\mathbb{I}_l$ is the identity
map in $\mathcal{M}_l$). A CP map is trace preserving if $\mathrm{Tr}
[\Phi(\rho)] = \mathrm{Tr} (\rho)$ for any $\rho \in \mathcal{M}_n$. Completely
positive trace preserving maps are referred to as Kraus maps; they represent the
reduced dynamics of quantum systems initially uncorrelated with the
environment~\cite{TarasovBook,Kraus83}.

Any Kraus map $\Phi$ can be represented using the operator-sum representation
(OSR) as $\Phi(\rho) = \sum_{i=1}^l K_i \rho K^{\dagger}_i$, where $\{ K_i \}$
is a set of complex $n \times n$ matrices satisfying the condition $\sum_{i=1}^l
K^{\dagger}_i K_i = \mathbb{I}_n$ to ensure trace
preservation~\cite{TarasovBook,Kraus83}. The OSR is not unique: any Kraus map
$\Phi$ can be represented using infinitely many sets of Kraus operators.

The all-to-one Kraus map for a given final state $\rho_{\rm f}=\sum_i
p_i|\phi_i\rangle\langle\phi_i|$ is defined as a Kraus map $\Phi_{\rho_{\rm f}}$
steering all initial states into $\rho_{\rm f}$, i.e., such that
$\Phi_{\rho_{\rm f}}(\rho)=\rho_{\rm f}$ for all $\rho\in{\cal
D}_n$~\cite{Wu07}. If $\rho_{\rm f}$ is a density matrix maximizing a given
objective $J$ and $c_*$ is a control producing the map $\Phi_{\rho_{\rm f}}$,
then this control will be simultaneously optimal for all initial states, i.e.,
the same $c_*$ will maximize the objective for any initial system state. An OSR
for a universally optimal Kraus map $\Phi_{\rho_{\rm f}}$ can be constructed by
using $K_{ij}=\sqrt{p_i}|\phi_i\rangle\langle\phi_j|$ as the Kraus operators.
Indeed, for any $\rho$, $\sum_{i,j=1}^n K_{ij}\rho K^\dagger_{ij}=\rho_{\rm f}$.

All-to-one Kraus maps were constructed theoretically in~\cite{Wu07}. The two
stage control scheme described in the present paper provides an approximate
physical realization of all-to-one Kraus maps $\Phi_{\rho_{\rm f}}$ for all
$\rho_{\rm f}$ (therefore any all-to-one Kraus map can be produced using this
scheme) for generic systems unitary controllable on the fast with respect to
$\tau_{\rm rel}$ time scale.

\section{Example: calcium atom} The scheme is illustrated below with an example
of a two-level atom whose all relevant parameters are known and it is easy to
analytically understand and visualize the controlled dynamics. We consider
calcium upper and lower levels $\rm 4^1 P$ and $\rm 4^1 S$ as two states
$|1\rangle$ and $|0\rangle$ of the two-level system. For this system the
transition frequency is $\om_{21}=4.5\times 10^{15}$~rad/s, the radiative
lifetime $t_{21}=4.5$~ns, the Einstein coefficient $A_{21}=1/t_{21}\approx
2.2\times 10^8$~s$^{-1}$, and the dipole moment $\mu_{12}=2.4\times 10^{-29}$
C$\cdot$m~\cite{Hilborn}. The method works equally well for any initial and
target states, we take for the sake of definiteness $\rho_0=|0\rangle\langle 0|$
and $\rho_{\rm f}=\frac{1}{4}|0\rangle\langle 0|+\frac{3}{4}|1\rangle \langle
1|$.

\begin{figure}[h]
\includegraphics[scale=0.45]{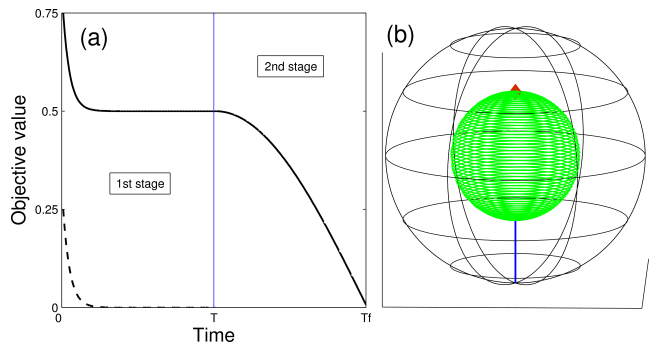}
\caption{(Color online). Application of the control scheme for engineering a
desired mixed state of two Ca energy levels. Subplot (a) shows the behavior of
the objective vs. time during the two stages of the control. The timescale is
not uniform, the duration of the first stage is $T=50$~ns and of the second
stage is $T_{\rm f}-T=1310$~fs. Solid line is for $\|\rho_t-\rho_{\rm f}\|$ and
dotted line for $\|\rho_t-\tilde\rho_{\rm f}\|$; the latter quantity almost
completely vanishes during the first stage of the control. Subplot (b) shows the
corresponding evolution of the Bloch vector. The initial system state
corresponds to the south pole, the vertical blue line represents incoherent
evolution during the first stage and the green line shows coherent rotations
during the second stage under the action of resonant field of amplitude
$E=10^7$~V/m. The evolved state approaches the target state indicated by the red
marker.}
\label{2level-1}
\end{figure}

\begin{figure}[h]
\includegraphics[scale=0.47]{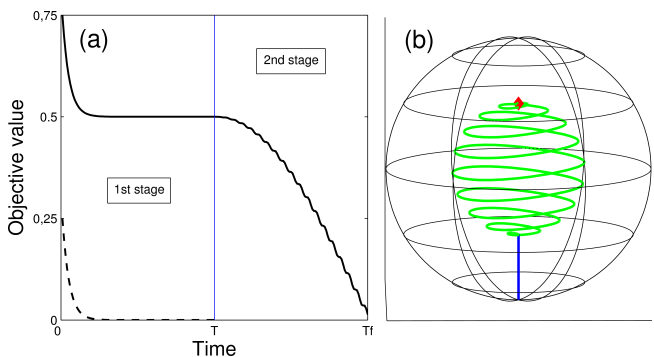}
\caption{(Color online). The only difference from the description of
Fig.~\ref{2level-1} is that $E=10^9$~V/m and $T_{\rm f}-T=13.1$~fs.}
\label{2level-2}
\end{figure}

The system is generic, all its relevant parameters are known and we can
analytically find optimal controls. The goal of the first (incoherent) stage is
to prepare the mixture $\tilde\rho_{\rm f}=\frac{3}{4}|0\rangle\langle
0|+\frac{1}{4}|1\rangle\langle 1|$. This goal is realized by applying to the
system incoherent radiation with spectral density satisfying $n_{\om_{21}}=1/2$
during time $T$ of several magnitudes of decoherence time $t_{21}$; we choose
$T=50$~ns. The goal of the second (coherent) stage is to rotate the state
produced at the end of the first stage to transform it into $\rho_{\rm f}$. This
goal can be realized for example by applying a resonant $\pi$-pulse
$E(t)=E\cos(\om_{12}t)$. Other methods of coherent control which produce the
same unitary transformation can be used as well. The electric field amplitude
which makes the Rabi frequency equal to the radiative decay rate is $E\approx
10^3$~V/m~\cite{Hilborn}. The proposed scheme requires the duration of the
second stage to be significantly shorter that the decay time. This can be
satisfied by choosing $E\gtrapprox 10^4$~V/m. We take resonant electromagnetic
field $E(t)=E\cos(\om_{12}t)$ of amplitude $E=10^7$~V/m acting on the system
during the time interval $T_{\rm f}-T=1310$~fs. The Rabi frequency for the field
of such amplitude is $\Omega_{\rm R}\approx 1/2320$~fs$^{-1}$, thus the field
acts as a $\pi$-pulse transforming the state $\tilde\rho_{\rm f}$ into
$\rho_{\rm f}$. The results of the numerical simulation are shown on
Fig.~\ref{2level-1}. The time interval $1310$~fs is much less than $t_{21}$ and
decoherence effects are negligible during the second stage. For illustrative
purpose of better visualization of the trajectory during the second stage we
provide in Fig.~\ref{2level-2} simulation results for $E=10^9$~V/m and $T_{\rm
f}-T=13.1$~fs. The method can be applied for producing arbitrary pure or mixed
target density matrices from any pure or mixed initial state thereby implying
complete density matrix controllability of this system.

\section*{Acknowledgments}
This research was supported by a Marie Curie International Incoming Fellowship
within the 7th European Community Framework Programme.

\end{document}